\documentclass[aps,prl,twocolumn,superscriptaddress,floatfix,showpacs]{revtex4}
\usepackage{graphicx}
\usepackage{amsmath}
\usepackage{amsfonts}
\usepackage{color}

\begin{document}

\title{Quantum transport signatures of chiral edge states in Sr$_2$RuO$_4$}

\author{Rakesh P. Tiwari}
\affiliation{
Department of Physics, University of Basel, Klingelbergstrasse 82, 
CH-4056 Basel, Switzerland}

\author{W. Belzig}
\affiliation{
Department of Physics, University of Konstanz, D-78457 Konstanz, Germany}

\author{Manfred Sigrist}
\affiliation{Theoretische Physik, ETH Z\"urich, CH-8093 Z\"urich, Switzerland}

\author{C. Bruder}
\affiliation{
Department of Physics, University of Basel, Klingelbergstrasse 82, 
CH-4056 Basel, Switzerland}

\begin{abstract}
  We investigate transport properties of a double quantum dot based
  Cooper pair splitter, where the superconducting lead consists of
  Sr$_2$RuO$_4$. The proposed device can be used to explore the
  symmetry of the superconducting order parameter in Sr$_2$RuO$_4$ by
  testing the presence of gapless chiral edge states, which are
  predicted to exist if the bulk superconductor is described by a
  chiral $p$--wave state. The odd orbital symmetry of the bulk order
  parameter ensures that we can realize a regime where the electrons
  tunneling into the double dot system come from the chiral edge
  states and thereby leave their signature in the conductance.  The
  proposed Cooper pair splitter has the potential to probe order
  parameters in unconventional superconductors.
\end{abstract}


\maketitle

\textit{Introduction\/}. -- Unconventional Cooper pairing features a
rich phenomenology in superconductivity ranging from non-standard
pairing mechanism to topologically non-trivial phases. An important
and much studies example is Sr$_2$RuO$_4$ which realizes most likely a
chiral $p$-wave phase, the quasi-two-dimensional analog of the A-phase
of superfluid $^3$He~\cite{mae94,Mackenzie-RMP,leg75}.  This phase has
topological character giving rise to chiral edge states (see for
example Ref.~\cite{fur01}).  Experimental evidence for chiral $p$-wave
pairing can be found in $\mu$SR measurements showing broken 
time-reversal symmetry~\cite{luk98}, Knight shift data demonstrating
in-plane equal-spin pairing~\cite{ish98}, and several more experiments
\cite{Mackenzie-RMP}. While edge states have been detected as
zero-bias anomalies in tunneling 
experiments~\cite{laube00,liu03,kashi11}, the search for the magnetic fields due
to the currents induced by the chiral edge states has turned out only negative 
results so far~\cite{tame03,kirtley07}. This discrepancy has recently
led to a renewed debate on the pairing symmetry realized in
Sr$_2$RuO$_4$~\cite{kallin12}.

In the present study we adopt the chiral $p$-wave symmetry for the
superconducting phase of Sr$_2$RuO$_4$ to study consequences in a
special quantum transport device.  The chiral $p$-wave phase is
characterized through pair wave function having a non-vanishing
angular momentum along the $z$-axis, $L_z=\pm 1$, and a spin-triplet
configuration with $S_z=0$ (in-plane equal-spin pairing).
In the standard {\bf d}-vector notation~\cite{leg75}, this order
parameter can be written as $\hat{\Delta} ({\bf k})=i{\bf d}({\bf
  k})\cdot{\boldsymbol{\sigma}}\sigma^{y}$, where $\sigma^{i}$
represent the Pauli matrices,
\begin{equation}
{\bf d}({\bf k})=\hat{z}\Delta \frac{k_x\pm ik_y}{k_F}\:,
\label{eq:dvector}
\end{equation}
and $k_F$ is the Fermi wave vector. Note that we use the coordinate
frame of the tetragonal crystal of Sr$_2$RuO$_4$ with $z$-axis
parallel to the four-fold crystalline symmetry axis.  Although
Sr$_2$RuO$_4$ is a three-band metal there are strong indications that
only one band, the genuinely two-dimensional $\gamma$-band, is
dominating superconductivity~\cite{huo13,wang13} so that it is
justified to use a single-band picture in the following. 

\begin{figure}
\includegraphics[width=.75\columnwidth,angle=0]{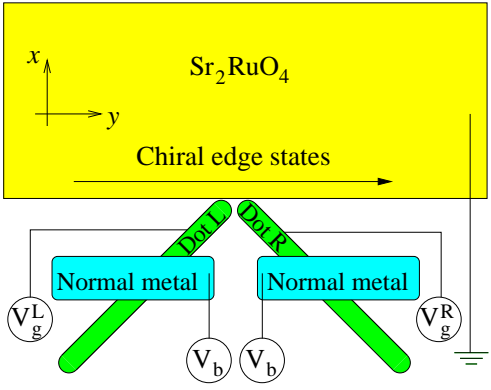}
\caption{Schematic of the device (not to scale).
Two single-walled carbon nanotubes (or one bent nanotube) are used to
form a double-dot system tunnel-coupled to a thin platelet of Sr$_2$RuO$_4$.
}
\label{fig:0}
\end{figure}

The topological index theorem necessitates the presence of gapless
chiral edge modes at the interface of such a chiral superconductor and
vacuum.  In this Letter we investigate the possibility of using
quantum transport measurements to directly probe these edge states. A
schematic view of the proposed device consisting of a double quantum
dot based Cooper pair splitter (CPS) is shown in Fig.~\ref{fig:0}. The
double dot (DD) system we consider is based on two single-walled
carbon nanotubes (or one bent nanotube), and the superconducting
electrode consists of a thin platelet of Sr$_2$RuO$_4$.  An SEM image
of an actual device employing a singlet superconductor is shown in
Fig. 1 of Ref.~\cite{schonenberger}.  Using a tunnel Hamiltonian
approach we calculate the coupling rates from the edge modes and the
bulk superconductor to the DD. We show that the subgap quantum
transport properties of such a DD tunnel-coupled to a Sr$_2$RuO$_4$
electrode and two normal leads reveal direct information about the
presence of chiral edge states in Sr$_2$RuO$_4$ in the presence of a
weak magnetic field.

\textit{Model Hamiltonian\/}. -- The total Hamiltonian of the system can
be written as
\begin{equation}
 \mathcal{H}=\mathcal{H}_{\text{DD}} + \mathcal{H}_{\text{NL}} +
\mathcal{H}_{\text{NR}} + \mathcal{H}_{\text{S}} + 
\mathcal{H}_{\text{SD}} + \mathcal{H}_{\text{ND}}\:,
\label{eq:totalham}
 \end{equation}
where 
\begin{eqnarray}
\mathcal{H}_{\text{DD}}&=&\sum_{i,\tau,\sigma}(\epsilon +
\tau\sigma\Delta_{\text{SO}})n_{i\tau\sigma}+\mathcal{H}_{\text{int}} 
\nonumber \\
&+& \Delta_{KK^{\prime}}\sum_{i,\sigma}(d^{\dagger}_{i\text{K}\sigma}
d_{i\text{K}^{\prime}\sigma}+d^{\dagger}_{i\text{K}^{\prime}\sigma}
d_{i\text{K}\sigma}) \nonumber\\
\label{eq:ddham}
\end{eqnarray}
describes the DD Hamiltonian~\cite{cottet:PRL} with $i\in\{\text{L},
\text{R}\}$ denoting the left and the right dot, and
$\sigma\in\{\uparrow,\downarrow\}$ (or equivalently $\sigma\in\{+,-\}$
in algebraic expressions) label the electronic spin states. In
addition, $\tau\in\{\text{K}, \text{K}^{\prime}\}$ (or equivalently
$\tau\in\{+,-\}$ in algebraic expressions) label the electronic
orbital states, a notation reminiscent of the valley degeneracy of
graphene. The term $\mathcal{H}_{\text{int}}$ accounts for the Coulomb
charging energy. We restrict ourselves to the regime with 
a single electron in each dot due to a strong intra-dot Coulomb
charging energy. The constant $\Delta_{\text{SO}}$ corresponds to an
effective spin-orbit coupling \cite{jesperson} and  the term
$\Delta_{\text{K}\text{K}^{\prime}}$ describes coupling between the K
and K$^{\prime}$ orbitals of dot $i$ due to disorder at the level of
the carbon nanotube atomic structure \cite{jesperson, liang, kuemmeth}.
Finally, $d^{\dagger}_{i\tau\sigma}$ denotes the DD creation operator
for an electron with spin $\sigma$ in orbital $\tau$ of dot $i$ and
$n_{i\tau\sigma}=d^{\dagger}_{i\tau\sigma}d_{i\tau\sigma}$, and
$\mathcal{H}_{\text{N}\eta}$ with $\eta\in\{\text{L}, \text{R}\}$
is the Hamiltonian of the left (L) and the right (R) normal leads
\begin{equation}
  \mathcal{H}_{\text{N}\eta}=
\sum_{k\sigma} \epsilon_{\eta k}c^{\dagger}_{\eta k \sigma}c_{\eta k \sigma}\:,
 \label{eq:nlham}
\end{equation}
where $k$ represents the orbital state in the normal leads and
$c^{\dagger}_{\eta k \sigma}$ denotes the normal lead electron
creation operator.

$\mathcal{H}_{\text{S}}$ is the Hamiltonian of the superconducting
lead. In our analysis we model the Sr$_2$RuO$_4$ as a thin platelet with a
two-dimensional spin-triplet mean-field Hamiltonian with ${\bf
  d}\parallel\hat{z}$, assuming that the energy band and the order
parameter have no momentum dependence in the $z$ direction. Thus
\begin{eqnarray}
\mathcal{H}_{\text{S}}&=& \sum_\sigma
\psi^\dagger_\sigma(\boldsymbol{r})h_0(\boldsymbol{r})
\psi_\sigma(\boldsymbol{r}) + \frac{1}{g}|
\boldsymbol{\eta}(\boldsymbol{r})|^2\nonumber\\
&-&\frac{i}{2k_F} \biggl\{\boldsymbol{\eta}(\boldsymbol{r})\cdot
\left[\psi^\dagger_\downarrow(\boldsymbol{r})\nabla
\psi^\dagger_\uparrow(\boldsymbol{r})+\psi^\dagger_\uparrow(\boldsymbol{r})
\nabla\psi^\dagger_\downarrow(\boldsymbol{r}) \right] \nonumber\\
&+&\boldsymbol{\eta}^*(\boldsymbol{r})\cdot\left[
\psi_\downarrow(\boldsymbol{r})\nabla\psi_\uparrow(\boldsymbol{r})
+\psi_\uparrow(\boldsymbol{r})\nabla\psi_\downarrow(\boldsymbol{r})
\right]
\biggr\}\:,
\label{MFHamiltonian}
\end{eqnarray}
where $h_0(\boldsymbol{r})=-\frac{\hbar^2}{2m}\nabla^2-\mu_{\text{S}}$,
$\psi_\sigma$ is the annihilation operator of an electron with spin
$\sigma$, $g$ is the coupling constant of the attractive interaction
that is responsible for $p$--wave pairing ($g>0$), and
$\nabla=(\partial_x,\partial_y)$.  The superconducting order parameter
$\boldsymbol{\eta}$ should satisfy the (self-consistency) gap
equation, obtained from
$(\delta/\delta\boldsymbol{\eta}^*)\langle
H_{\text{S}}\rangle=0$~\cite{fur01},
\begin{equation}
\boldsymbol{\eta}=(\eta_x,\eta_y)=
\frac{ig}{2k_F}
\left\langle\psi_\uparrow(\boldsymbol{r})[\nabla
\psi_\downarrow(\boldsymbol{r})] - [\nabla\psi_\uparrow(\boldsymbol{r})] 
\psi_\downarrow(\boldsymbol{r})\right\rangle.
\label{selfconsistent}
\end{equation}
For a uniform system with a $\boldsymbol{d}$-vector corresponding to
Eq.~(\ref{eq:dvector}), this implies $\boldsymbol{\eta}=\Delta(1,\pm
i)$. Within this model a description of the chiral edge states can be
obtained by solving the Bogoliubov--de Gennes (BdG) equation across
the boundary of the superconductor, assuming a simplified gap function
$\boldsymbol{\eta}=\Theta(x)\Delta(1,\pm i)$, where $\Theta(x)$ is the
Heaviside step function~\cite{foot1}. Here, the $x$ direction is
assumed to be normal to the boundary of the superconductor.  
The field operators $\psi_\sigma$ can be expanded in terms of Bogoliubov
operators as~\cite{fur01}
\begin{equation}
\left(\begin{array}{c}
 \psi_\uparrow(\bf{r}) \\
 \psi_\downarrow^\dagger(\bf{r})
\end{array}\right)
=\sum_{{\bf k}}
\left(\begin{array}{cc}
       u_{\bf{k}} & v_{\bf{k}}^{\ast} \\
       v_{\bf{k}} & u_{\bf{k}}^{\ast}
      \end{array}
\right)
\left(\begin{array}{c}
       \gamma_{{\bf{k}}\uparrow} \\
       \gamma_{{\bf{k}}\downarrow}^\dagger
      \end{array}
\right)\:,
\label{eq:expansion}
\end{equation}
where the Bogoliubov operators are defined as $\gamma_{\bf{k}\sigma}|0
\rangle=0$, with $|0 \rangle$ denoting the ground state of the
superconductor. The bound-state solutions of the BdG equations
satisfying the boundary condition $u_{{\bf k}}=v_{{\bf k}}=0$ at $x=0$
denote the chiral edge states, and are given by
\begin{equation}
\left(\begin{array}{c}
       u_{\bf{k}}(\bf{r}) \\
       v_{\bf{k}}(\bf{r})
      \end{array}
\right)=\mathcal{N}
\exp\left(-\frac{x}{\xi_0}+i k_y y\right)
\sin(k_x x)
\left(\begin{array}{c}
       e^{i\pi/4} \\
       e^{-i\pi/4}
      \end{array}
\right)\:,
\label{eq:boundstate}
\end{equation}
with energy eigenvalues $\epsilon_\chi(k_y)=\Delta\frac{k_y}{k_F}$ 
and the normalization factor
$\mathcal{N}=\sqrt{\frac{2}{\xi_0L_y}}$. Here $L_y$ is the length of
the superconductor in $y$ direction. These edge state solutions
decaying in the bulk on the length scale $\xi_0=\hbar v_F/\Delta$ have
been obtained within the Andreev approximation ($\Delta\ll\mu_S$).

The tunnel Hamiltonian between dot $l\in\{L,R\}$ and the
point ${\bf r}_l$ in the superconductor is 
\begin{equation}
\mathcal{H}_{\text{SD}}=\sum_{l,\tau}T_{\text{SD}}[d^{\dagger}_{l\tau\uparrow}
\psi_{\uparrow}({\bf r}_l) + 
d^{\dagger}_{l\bar{\tau}\downarrow}\psi_{\downarrow}({\bf r}_l)]+ h.c.\:,
\label{eq:hsd}
\end{equation}
here $\psi_{\sigma}({\bf r}_l)$ annihilates an electron with spin
$\sigma$ at site ${\bf r}_l$, and $d^{\dagger}_{l\tau\sigma}$ creates
it again (with the same spin) at dot $l$ and in electronic orbital
$\tau$ with amplitude $T_{SD}$. Finally,
\begin{equation}
\mathcal{H}_{ND}=\sum_{k,\tau,\sigma}[T_{\text{NL}}c^{\dagger}_{Lk\sigma}
d_{L\tau\sigma} + T_{\text{NR}}c^{\dagger}_{Rk\sigma}d_{R\tau\sigma}]+ h.c.
\label{eq:hnd}
\end{equation}
describes the tunnel coupling between the left (right) dot and 
the left (right) normal lead with amplitude $T_{\text{NL}}$ and
$T_{\text{NR}}$ respectively.

\textit{Tunneling between the superconductor and the DD\/}. -- 
In our device the coherent injection of electrons in the DD via tunneling
appears in the following order. A Cooper pair breaks up in Sr$_2$RuO$_4$,
one electron with spin $\sigma$ tunnels to one of the dots from the
point of the superconductor nearest to this dot. This results in a
virtual state where the other electron either creates a quasiparticle
with energy $E_{\bf k} > \Delta$ (referred to as Case I below) or it
occupies an empty edge state with energy $\epsilon_\chi(k_y)$ (referred to
as Case II below). The second electron with spin $-\sigma$ then
tunnels to the other empty dot before the first electron with spin
$\sigma$ tunnels out to the normal lead making the tunneling of both
electrons almost simultaneous (within the uncertainty time
$\hbar/\Delta$). Reference~\cite{recher} describes such a device for
singlet superconductors.

To elucidate the role of chiral edge states in the Cooper pair
tunneling described above, we derive the effective Hamiltonian of the
DD--normal lead system by integrating out the superconductor's degrees
of freedom and including terms up to second order in
$T_{\text{SD}}$~\cite{recher}. Writing
$\mathcal{H}=\mathcal{H}_0+\mathcal{H}_{\text{SD}}$ we obtain
$\mathcal{H}_{\text{eff}}=\mathcal{H}_{\text{DD}} +
\mathcal{H}_{\text{NL}} +\mathcal{H}_{\text{NR}} -\mu_S +
\mathcal{H}^{\text{eff}}_{\text{SD}} +\mathcal{H}_{\text{ND}}$, where

\begin{equation}
\mathcal{H}_{\text{SD}}^{\text{eff}}=\lim_{\zeta\rightarrow0^{+}} \langle
0\mid\mathcal{H}_{SD}\frac{1}{i\zeta-\mathcal{H}_0}\mathcal{H}_{SD}
\mid0 \rangle\:.
\label{eq:hsdeff}
\end{equation}
We find that 
\begin{equation}
\mathcal{H}_{\text{SD}}^{\text{eff}}=T_{SD}^2\sum_{\tau}t_{eh}(d^{\dagger}_{\text{R}
\tau\uparrow}d^{\dagger}_{\text{L}\overline{\tau}\downarrow}+
d^{\dagger}_{\text{R}\overline{\tau}\downarrow}d^{\dagger}_{\text{L}
\tau\uparrow}) + h.c.\:, 
\label{eq:hsdeffresult}
\end{equation}

where $t_{eh}=t_{eh}^{\text{I}} + t_{eh}^{\text{II}}$ is the effective
amplitude for the tunnel coupling between the superconductor and the
DD system, including contributions from both the quasiparticle states
denoted by $t_{eh}^{\text{I}}$ and the chiral edge states denoted by
$t_{eh}^{\text{II}}$. Within the BdG theory (and using that for a
spin-triplet superconductor $u_{{\bf k}}v_{{\bf k}}=-u_{-{\bf
    k}}v_{-{\bf k}}$) the quasiparticle contribution is
\begin{equation}
 t_{eh}^{\text{I}}=4i\sum_{\bf k}
\frac{u_{\bf k}v_{\bf k}}{E_{\bf k}}\sin{({\bf k}\cdot\delta{\bf r})}\:,
 \label{eq:teh1}
\end{equation}
where $\delta{\bf r}={\bf r}_R - {\bf r}_L$ and 
$E_{{\bf k}}^2=\Delta^2+\xi_{{\bf k}}^2$ denotes the quasiparticle
spectrum. The sum over ${\bf k}$ can be performed by linearizing the
spectrum around the Fermi level with Fermi wave vector $k_F$. Finally
we obtain
\begin{equation}
 t_{eh}^{\text{I}}=i 2\pi \rho_S J_1(k_F\delta r)\:,
 \label{eq:teh1final}
\end{equation}
where $\rho_S$ is the (normal-state) density of states of the
superconductor at the Fermi level, $J_1(k_F\delta r)$ denotes the
first-order Bessel function and $\delta r=|{\bf r}_R-{\bf r}_L|$.
Note that $\lim_{\delta r \to 0} t_{eh}^{\text{I}}=0$.

For the chiral edge states the Bogoliubov transformation is given by
Eq.~(\ref{eq:boundstate}). A similar calculation yields
\begin{equation}
t_{eh}^{\text{II}}=2\lim_{\zeta\rightarrow0^{+}}\sum_{{\bf k}}\frac{u_{{\bf k}}({\bf
    r}_R)v^{\ast}_{{\bf k}}({\bf r}_{L})}{\epsilon_\chi(k_y)+ \epsilon-i\zeta}\:,
\label{eq:teh2}
\end{equation}
where $\epsilon$ indicates the double-dot orbital energy, see
Eq.~(\ref{eq:ddham}), and the sum over ${\bf k}$ represents a
summation over one dimension momentum along the edge.  Interestingly
we find that this contribution remains finite even when ${\bf
  r}_L={\bf r}_R$. As we are interested in probing the chiral edge
states, we consider a CPS device where the \textit{two} electrons
tunneling into the DD system satisfy ${\bf r}_L\approx{\bf r}_R$
(implying that
$t_{eh}^{\text{II}}/t_{eh}^{\text{I}}\rightarrow\infty$).  Therefore
in our device the electron tunneling amplitude between the
superconductor and the DD system contains only the edge state
contribution $t_{eh}\approx t_{eh}^{\text{II}}$.

\textit{Tunneling between the normal leads and the DD\/}. -- In the
sequential-tunneling regime a convenient description of quantum
transport in our device can be obtained within the Master equation
formalism. By integrating out the normal-lead degrees of freedom, the
tunneling of an electron between \textit{one} quantum dot and the
corresponding normal lead is described by a tunnel rate
$\Gamma_{\text{N}\eta}=2\pi\rho_\eta |T_{\text{N}\eta}|^2$, where
$\eta\in\{L,R\}$ denotes the left (L) and the right (L) lead, and
$\rho_{\text{N}\eta}$ denotes the corresponding density of states
(assumed to be be constant in the energy window relevant for
transport).  For simplicity we assume
$\Gamma_{\text{N}L}=\Gamma_{\text{N}R}=\Gamma_{\text{N}}$. In the
master-equation description of the dynamics of our device, to
lowest order in $\Gamma_N$ we only need to consider the occupation
probabilities of the eigenstates of the effective Hamiltonian
$\mathcal{H}_{\text{eff}}^{\text{DD}}=
\mathcal{H}_{\text{DD}}+\mathcal{H}_{\text{SD}}^{\text{eff}}$.

Due to Coulomb blockade, the DD eigenstates participating in transport
are the double-dot empty state $|0,0\rangle$, the singly occupied
states $|\tau\sigma,0\rangle=d^{\dagger}_{L\tau\sigma}|0,0\rangle$ and
$|0,\tau\sigma\rangle=d^{\dagger}_{R\tau\sigma}|0,0\rangle$, and the
nonlocal doubly occupied states
$|\tau\sigma,\tau^{\prime}\sigma^{\prime}\rangle=
d^{\dagger}_{L\tau\sigma}d^{\dagger}_{R\tau^{\prime}
  \sigma^{\prime}}|0,0\rangle$. For $t_{eh}=0$,
$\Delta_{\text{SO}}=0$, and $\Delta_{KK^\prime}=0$, to create the
doubly-occupied states of $\mathcal{H}_{\text{eff}}^{\text{DD}}$ will
cost an energy $2\epsilon$. In our device we assume
$t_{eh}\ll\{\Delta_{SO}, \Delta_{KK^{\prime}}\}$, this allows us to
isolate a subset of \text{five} lowest-energy eigenstates of
$\mathcal{H}_{\text{eff}}^{\text{DD}}$ in the regime $\epsilon\sim
\Delta_r$, which are separated from all other doubly-occupied
eigenstates by at least $\sim 2\Delta_r$, where
$\Delta_r=\sqrt{\Delta_{\text{SO}}^2 +\Delta_{KK^{\prime}}^2}$.  A
similar calculation for a singlet superconductor is presented in
Ref.~\cite{cottet:PRL} and its supplemental information.  These
\textit{five} states include two triplet states
$|T_{\uparrow\uparrow}\rangle$, $|T_{\downarrow\downarrow}\rangle$ as
well as the singlet state $|S_{\uparrow\downarrow}\rangle$ with
energies $E=2\epsilon -2\Delta_r$, and the two hybridized states
$|V_n\rangle=\sqrt{1-|v_n|^2}|0,0\rangle +
v_n|T_{\uparrow\downarrow}\rangle$ with energies
$E_n=(\epsilon-\Delta_r-(-1)^n\sqrt{2|t_{eh}|^2+
  (\epsilon-\Delta_r)^2})$ for $n\in\{1,2\}$. Here,
$v_1=t_{eh}/\sqrt{2|t_{eh}|^2+ (\epsilon-\Delta_r)E_2}$ and
$v_2=t_{eh}/\sqrt{2|t_{eh}|^2+ (\epsilon-\Delta_r)E_1}$.  In our
device the coherent injection of Cooper pairs with symmetry as
described by Eq.~(\ref{eq:dvector}) ensures that the triplet states
$|T_{\uparrow\uparrow}\rangle$ and $|T_{\downarrow\downarrow}\rangle$,
as well as the singlet state $|S_{\uparrow\downarrow}\rangle$, are not
populated.
\begin{figure}
\includegraphics[width=.77\columnwidth,angle=270]{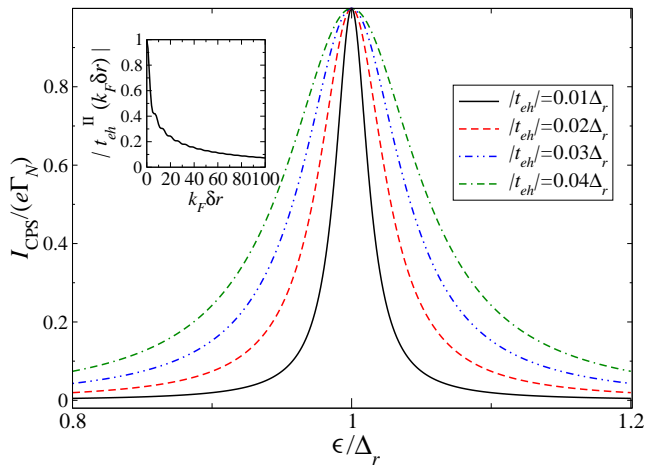}
\caption{$I_{\text{CPS}}$ as a function of $\epsilon$ 
that is tunable
  by (symmetric) external gate voltages, calculated using
  Eq.~(\ref{eq:icps}) for various values of $|t_{eh}|$. Inset: 
$|t_{eh}^{\text{II}}(k_F\delta r)|/|t_{eh}^{\text{II}}(0)|$ as a function of $k_F\delta r$ for
$\epsilon/\Delta=0.01$ calculated using Eq.~(\ref{eq:teh2}). 
}
\label{fig:1}
\end{figure}
The occupation probabilities $P_{V_1}$, $P_{V_2}$ and
$P_{\text{single}}$ of the states $|V_1\rangle$, $|V_2\rangle$, and the
global probability of having a singly-occupied DD state, respectively
satisfy a master equation $dP/dt = M P$, where
$P=\left[P_{V_1}, P_{V_2}, P_{\text{single}}\right]^T$ and
\begin{equation}
 M=\Gamma_N\left[
 \begin{array}{ccc}
  -2|v_1|^2 & 0  & 1-|v_1|^2 \\
  0 & -2|v_2|^2 & 1-|v_2|^2 \\
  2|v_1|^2 & 2|v_2|^2 & -1
 \end{array}
 \right]\:.
 \label{eq:mmatrix}
\end{equation}
Here we have assumed that the applied bias voltage $V$ between the
superconductor and the normal leads resulting in the subgap transport
is such that the single electrons can tunnel from the DD system to the
normal leads but not vice-versa. By solving the equation
$MP_{\text{stat}}=0$, we can calculate the steady-state value of the
occupation probabilities $P_{\text{stat}}$ which allows us to compute
the dc current response of our device as $I_{\text{CPS}}=R\cdot
P_{\text{stat}}$, where $R=e\Gamma_N\left[2|v_1|^2, 2|v_2|^2,
  1\right]$. This current response can be understood in terms 
of state cycles which produce a flow of electrons towards
the normal leads~\cite{cottet:PRL}. Due to the injection of Cooper
pairs the DD system starts in a state $|V_n\rangle$, where
$n\in\{1,2\}$. Now a single-electron tunneling event to either of the
leads results in a singly-occupied state in the DD system, another
such tunneling event can then cause a transition back to $|V_n\rangle$
because $|V_n\rangle$ has a $|0,0\rangle$ component. We find
\begin{align}
I_{\text{CPS}}&=e\Gamma_N\frac{2|t_{eh}|^2}{2|t_{eh}|^2 
+ (\epsilon-\Delta_r)^2}\nonumber \\
&=e\Gamma_N\frac{2(|t_{eh}^{\text{I}}|^2+|t_{eh}^{\text{II}}|^2)}
{2(|t_{eh}^{\text{I}}|^2+|t_{eh}^{\text{II}}|^2)+ (\epsilon-\Delta_r)^2}
 \:.
\label{eq:icps}
\end{align}
Figure~\ref{fig:1} shows $I_{\text{CPS}}$ for
various values of $t_{eh}$. In the limit $\delta r \to 0$, i.e.,
$t_{eh}\approx t_{eh}^{\text{II}}$, the current given in
Eq.~(\ref{eq:icps}) is solely due to the chiral edge states. Even if
this condition is not satisfied, our device can provide a signature of
the presence of the chiral edge states. By measuring the full width at
half maximum (FWHM), we can extract information about the dependence of 
$t_{eh}^{\text{I}}$ and $t_{eh}^{\text{II}}$ on various parameters of
the device like e.g. $k_F\delta r$.

\textit{Detecting chirality\/}. Our measuring scheme so far does not
allow to distinguish chiral from non-chiral edge states. We may,
however, run a small in-plane supercurrent along the edge of the
superconductor (due to Meissner screening it will be concentrated
within the London penetration depth). The effect of the current is to
increase or decrease the density of states of the edge states
depending on the chirality or the direction of the current,
respectively~\cite{yok08}.  As a consequence $t_{eh}^{\text{II}}$ and
thereby the FWHM of the current shown in Fig.~\ref{fig:1} would depend
essentially linearly on the supercurrent, providing a clear signature
of chirality. This would make it even possible to determine the sign of
chirality of the superconducting phase for a given setup.

\textit{Conclusions\/}. We have calculated the current response of a
double quantum dot based Cooper pair splitter, where the
superconducting electrode is a thin platelet of Sr$_2$RuO$_4$.  The
expression for the current response provides direct evidence for the
chiral edge states predicted to exist along the boundaries of the
superconductor. By applying a small supercurrent along the edge of the
superconducting platelet our device can also detect the sign of the
chirality. An experimental realization of this device would help to
settle the debate on the order parameter of superconducting Sr$_2$RuO$_4$.

\textit{Acknowledgments\/}. RPT and CB acknowledge financial support
by the Swiss SNF and the NCCR Quantum Science and Technology. 
WB acknowledges financial support from the DFG through BE3803/3 and the Baden 
W\"urttemberg Stiftung.

\end{document}